\documentclass{jpconf}
\usepackage{amsmath}
\usepackage{epsfig}
\begin{document}
\title{Systematic Properties of the Tsallis Distribution:\\
Energy Dependence of Parameters.}
%
% subtitle is optional
\author{J. Cleymans}
\address{UCT-CERN Research Centre and Department of Physics\\
University of Cape Town,
Rondebosch, South Africa 
	  }

\ead{jean.cleymans@uct.ac.za}
\begin{abstract}
Changes in the transverse momentum distributions with beam energy
are studied using the Tsallis distribution as a parameterization.
The dependence of the Tsallis parameters $q$, $T$ and  the volume  
  on beam energy is determined.
The Tsallis parameter q shows a weak but clear increase with beam energy
with the highest value being approximately 1.15.
The Tsallis temperature and volume are consistent with being independent 
of beam energy within experimental uncertainties.
\end{abstract}
In the analysis of new data, a Tsallis-like distribution 
gives excellent fits to the transverse momentum 
distributions  as shown by the  
 STAR~\cite{STAR} and PHENIX~\cite{PHENIX} collaborations at RHIC and by the  
ALICE~\cite{ALICE}, ATLAS~\cite{ATLAS} and CMS~\cite{CMS} collaborations at 
the LHC.
In this talk we review the parameterization used by these groups and propose a slightly different one
which leads to a more consistent interpretation and has the bonus of being  
thermodynamically consistent~\cite{parvan}.

In the framework of Tsallis 
statistics~\cite{tsallis1,tsallis2,miller,worku1,worku2} the entropy, $S$,
the particle number, $N$,  the energy $E$ and the pressure $P$ 
are given by corresponding
integrals over the  Tsallis distribution: 
\begin{equation}
f \equiv  \left[ 1 + (q-1) \frac{E-\mu}{T}\right]^{-\frac{1}{q-1}} .\label{tsallis} 
\end{equation}
where $T$ and $\mu$ are the temperature and the chemical potential, $q$ is a new variable 
often refereed to as Tsallis parameter. The relevant thermodynamic quantities are 
given by (see e.g.~\cite{worku2}
\begin{eqnarray}
S &=& -gV\int\frac{d^3p}{(2\pi)^3}\left[ f^{q}\ln_{q} f -f \right],\label{entropy} \\
N &=&  gV\int\frac{d^3p}{(2\pi)^3} f^q ,\label{number} \\
	E &=&  gV\int\frac{d^3p}{(2\pi)^3}\sqrt{p^2+m^2} f^q ,\label{epsilon}\\
P &=&  g\int\frac{d^3p}{(2\pi)^3}\frac{p^2}{3E} f^q\label{pressure} .
\end{eqnarray}
$V$ is the volume,  $g$ is the degeneracy factor.  
The short-hand notation
\begin{equation}
\ln_q (x)\equiv \frac{x^{1-q}-1}{1-q} , \label{suba} 
\end{equation}
often referred to as q-logarithm has been used.
These expressions are  thermodynamically consistent, 
e.g. it can be shown~\cite{worku2} that consistency relations
of the type
\begin{equation}
 N = V\left.\frac{\partial P}{\partial \mu}\right|_{T},\label{consistency}
\end{equation}
and 
\begin{equation}
	\epsilon + P = Ts + \mu n
\end{equation}
(where $n, s, \epsilon$ refer to the densities of the corresponding quantities)
are indeed satisfied.

Following from Eq.~(\ref{number}), the momentum distribution is given by,
\begin{equation}
\frac{d^{3}N}{d^3p} = 
\frac{gV}{(2\pi)^3}
\left[1+(q-1)\frac{E -\mu}{T}\right]^{-q/(q-1)},
\label{tsallismu}
\end{equation}
or, expressed in terms of transverse momentum, $p_T$,  
 transverse mass, $m_T \equiv \sqrt{p_T^2+ m ^2}$, and rapidity  $y$  
\begin{equation}
\frac{d^{2}N}{dp_T~dy} = 
gV\frac{p_Tm_T\cosh y}{(2\pi)^2}
\left[1+(q-1)\frac{m_T\cosh y -\mu}{T}\right]^{-q/(q-1)} .
\label{tsallismu1}
\end{equation}
At mid-rapidity, $y = 0$, and for zero chemical potential, as is relevant at 
the LHC, this reduces to 
\begin{equation}
\left.\frac{d^{2}N}{dp_T~dy}\right|_{y=0} = 
gV\frac{p_Tm_T}{(2\pi)^2}
\left[1+(q-1)\frac{m_T}{T}\right]^{-q/(q-1)}.
\label{tsallisfit1}
\end{equation}
In the limit where the parameter $q$ goes to 1 it is well-known that this reduces 
the standard Boltzmann distribution.
The parameterization given in Eq.~(\ref{tsallismu1}) is close to
the one used in~\cite{STAR,PHENIX,ALICE,ATLAS,CMS}: 
\begin{equation}
  \frac{d^2N}{dp_T\,dy} = p_T \frac{dN}
  {dy} \frac{(n-1)(n-2)}{nC(nC + m_{0} (n-2))}
 \left[ 1 + \frac{m_T - m_{0}}{nC} \right]^{-n}  ,
\label{alice}
\end{equation}
where $n$ (not to be confused with the particle density) and $C$ are fit parameters and 
has been discussed in detail in Ref.~\cite{parvan}.
%This corresponds to substituting 
%\begin{equation}
%n\rightarrow \frac{q}{q-1}   ,
%\label{n}
%\end{equation}
%and 
%\begin{equation}
%nC  \rightarrow \frac{T+m_0(q-1)}{q-1}  .
%\label{nC}
%\end{equation}
%After  this substitution Eq.~\ref{alice} becomes
%\begin{eqnarray}
%  \frac{d^2N}{dp_T\,dy} =&& p_{T} \frac{{\rm d}N}
%  {{\rm d}y} \frac{(n-1)(n-2)}{nC(nC + m_{0} (n-2))}\nonumber\\ 
%&&\left[\frac{T}{T+m_0(q-1)}\right]^{-q/(q-1)}\nonumber\\
%&&\left[ 1 + (q-1)\frac{m_T}{T} \right]^{-q/(q-1)}  .
%\label{alice2}
%\end{eqnarray}
%%
%
At mid-rapidity $y=0$ and zero chemical potential,
they have a similar dependence on the 
transverse momentum as Eq.~(\ref{tsallisfit1}) 
apart from an additional  factor $m_T$ on the right-hand side. % on the right-hand side of Eq.~(\ref{tsallismu1}).
However, the inclusion of the rest mass% in the substitution Eq.~(\ref{nC})
is not in agreement with the Tsallis distribution as it breaks 
$m_T$ scaling which is present in Eq.~(\ref{tsallismu1})
but not in Eq.~(\ref{alice}). 
The inclusion of the factor $m_T$ 
leads to a more consistent interpretation of the variables $q$ and $T$.
\\

A very good description of transverse momenta distributions at RHIC has also been
obtained in Refs~\cite{coalesce1,coalesce2} on the basis of a coalescence model 
where the Tsallis distribution is used for quarks. 
Tsallis fits have also been considered in Ref.~\cite{wong,wibig1,wibig2} but 
with a different power law leading to smaller values of the Tsallis parameter  $q$.

Interesting results were obtained in 
Refs.~\cite{deppman,deppman3}
where spectra for identified particles were analyzed and the resulting
values for the parameters $q$ and $T$ were considered.\\
It has been shown that, at lower energies it is possible to describe
transverse momentum spectra in $p-p$ collisions with Boltzmann distributions 
provided one takes into account
the decays of resonances~\cite{becattini}. 
This is not in contradiction with the results presented here as the 
transverse momentum distributions become more and more Boltzmann like at lower energies
as can be seen in the values of $q$ which are closer to one as the energy is decreased.

The energy dependence in $p-p$ collisions can be determined by 
studying data for charged particles  at beam energies
of 0.54~\cite{UA1}, 0.9, 2.36  and 7 TeV~\cite{ALICE,ATLAS,CMS}. 
These involve distributions summed over charged particles.
The fits were performed using a sum of three Tsallis distributions, the 
first one for $\pi^+$, the
second one for $K^+$ and the third one for protons $p$. The relative 
weights between these were 
simply determined by the corresponding degeneracy factors, i.e. 1 for for $\pi^+$ and $K^+$
and 2 for  protons. 
The fit was taken at mid-rapidity and for $\mu = 0$ using the following expression:
\begin{equation}\label{tsallisfit}
   \left.  \frac{1}{2\pi p_{T}} \frac{d^{2}N(\mathrm{charged \
particles})}{dp_{T}dy}\right|_{y=0} = \frac{2V}{(2\pi)^{3}}
\sum\limits_{i=1}^{3} g_{i} m_{T,i}
\left[1+(q-1)\frac{m_{T,i}}{T}\right]^{-\frac{q}{q-1}},
\end{equation}
where $i=(\pi^{+},K^{+},p)$  and
$g_{\pi^{+}}=1$, $g_{K^{+}}=1$ and $g_{p}=2$. The factor $2$ in front
of the right hand side of this equation takes into account the
contribution of the antiparticles $(\pi^{-},K^{-},\bar{p})$. 
A comparison with charged particle distributions is shown in Fig.~1.
It is to be noted that the Tsallis distribution presented above also gives an excellent description of 
transverse momentum distributions in $p-Pb$ collisions  at all pseudorapidity intervals
obtained by the ALICE collaboration~\cite{ALICEpPb}.
The Tsallis parameters $q$ and $T$ needed to describe the transverse momentum
distributions of charged particles are shown in Fig.~2. The value of $q$t has a tendency to increase slowly
with increasing  energy~\cite{parvan} while no clear energy dependence for $T$ can be discerned.\\
\begin{figure}[tbh]
\begin{center}
\includegraphics[width=0.9\linewidth,height=9cm]{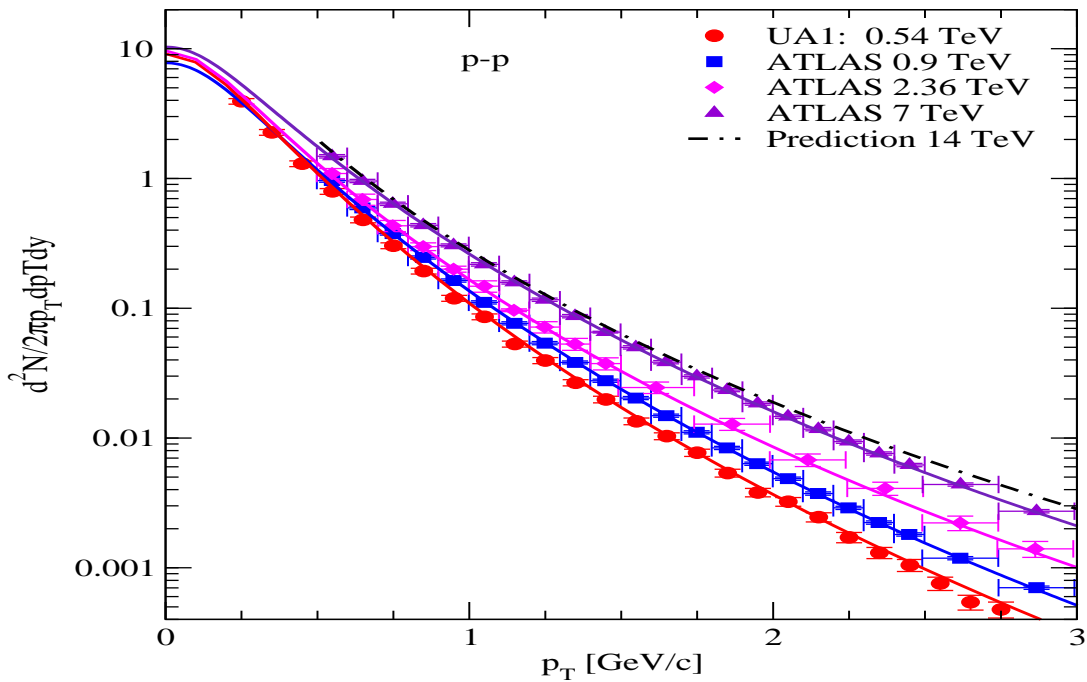}
\label{atlas}
\caption{Fits to transverse momentum distributions of charged  particles~\cite{ATLAS,UA1} using the Tsallis distribution.}
\end{center}
\end{figure}
%%%%%%%%%%%%%%%%%%%%%%%%%%%%%%%%%%%%%%%%%%%%%%%%%%%%%%%%%%%
\begin{figure}[tbh]
\begin{center}
\includegraphics[width=0.8\linewidth,height=8cm]{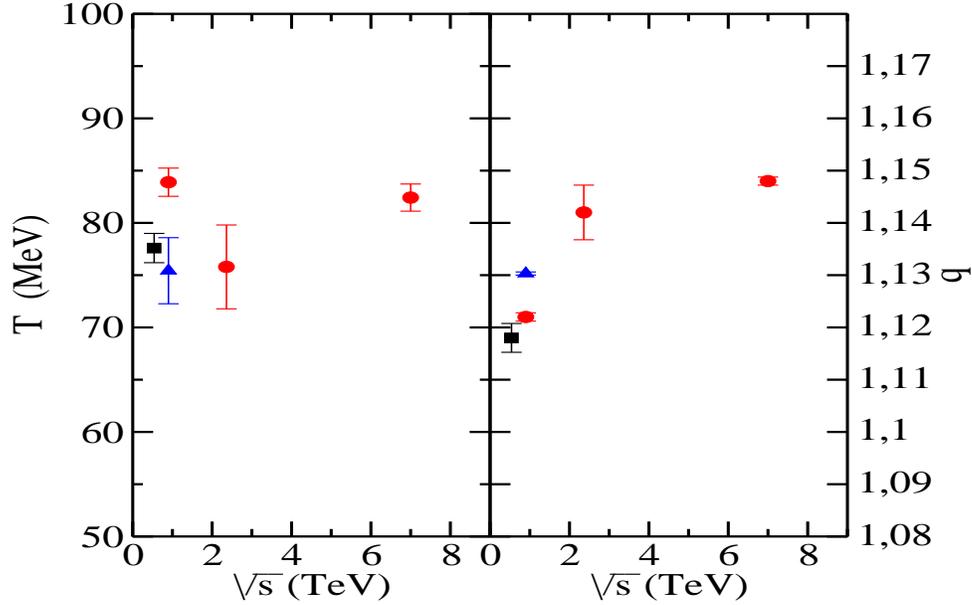}
\label{azmi_T}
\caption{Values of the Tsallis parameters $T$ (left) and $q$ (right),
as a function of beam energy, obtained from fits
to the transverse momentum distributions of charged particles as described in the text.
The square point is from~\cite{UA1}, the round points are from~\cite{ATLAS} 
while the triangle point is from~\cite{ALICE}.} 
\end{center}
\end{figure}
%
%%%%%%%%%%%%%%%%%%%%%%%%%%%%%%%%%%%%%%%%%%%%%%%%%%%%%%%%%%%
%
In conclusion, the Tsallis distribution 
leads to excellent fits to the transverse momentum distributions
in high energy $p-p$ and $p-Pb$ collisions. 
By comparing results from UA1~\cite{UA1} to results obtained at the 
LHC~\cite{ATLAS,CMS} it has been possible to extract the parameters
$q,T$ and $V$ for a wide range of energies~\cite{parvan}.
A consistent picture emerges from a comparison of fits using the Tsallis
distribution in high energy $p-p$ collisions.

\section*{References}


\begin{thebibliography}{90}
	\bibitem{STAR} B. I. Abelev et al. (STAR  Collaboration), Phys. Rev. C {\textbf{75}}, 064901 (2007).
	\bibitem{PHENIX} A. Adare et al. (PHENIX Collaboration), Phys. Rev. C {\textbf{83}}, 052004, (2010); Phys. Rev. C {\textbf{83}}, 064903 (2011).
\bibitem{ALICE} K. Aamodt, et al. (ALICE Collaboration), 
	Eur. Phys. J. C {\textbf{71}} 1655 (2011); Phys. Lett. B {\textbf{693}} (2010) 53.
\bibitem{ATLAS} G. Aad, et al. (ATLAS Collaboration), 
New J. Phys. {\textbf{13}} (2011) 053033.
\bibitem{CMS} V. Khachatryan, et al. (CMS Collaboration), Phys. Rev. Lett. {\textbf{105}} (2010) 022002.
\bibitem{tsallis1} C. Tsallis, J. Statist. Phys. {\bf 52}, 479 (1988).
%
\bibitem{parvan}
  J.~Cleymans, G.I.~Lykasov, A.S.~Parvan, A.S.~Sorin, O.V.~Teryaev, D. Worku
  Phys.\ Lett. B {\bf 723} (2013) 351.
%
\bibitem{tsallis2} C. Tsallis, R. S. Mendes, A. R. Plastino, Physica A {\bf 261}, 534 (1998).
\bibitem{miller} J. M. Conroy, H. G. Miller, A. R. Plastino, 
Phys. Lett. A {\bf 374}, 4581 (2010).
\bibitem{worku1} J. Cleymans and D. Worku, J. Phys. G {\bf 39} (2012) 025006.
\bibitem{worku2} J. Cleymans and D. Worku, 
Eur. Phys. J. A {\bf 48} (2012) 160. 
\bibitem{coalesce1} K. \"Urm\"ossy, T.S. Bir\'{o}, Phys. Lett. B {\textbf{689}} (2010)  14.
\bibitem{coalesce2}  K. \"Urm\"ossy, T.S. Bir\'{o}, J. Phys. G {\textbf{36}} (2009)  064044.
%
%
\bibitem{wong} Cheuk-Yin Wong, G. Wilk, Acta Physica Polonica, {\textbf{43}} (2012) 2047.
\bibitem{wibig1} T. Wibig,  J. Phys. G: Nucl. Part. Phys. {\textbf{37}} 115009 (2010).
\bibitem{wibig2} T. Wibig, I. Kurp, JHEP {\textbf{0312}} 039 (2003).
%
\bibitem{deppman} L. Marques, E.Andrade-II, A. Deppman, arXiv:1210.1725[hep-ph]
%
%
\bibitem{deppman3} I. Sena, A. Deppman, Eur. Phys. J. A 49 (2013) 17; arXiv:1209.2367[hep-ph]
%
%
%\bibitem{azmi} M.D.~Azmi and J. Cleymans (in preparation).
%
%
\bibitem{becattini} F. Becattini and G. Passaleva, Eur. Phys.. J. C {\bf 23} (2002) 551.
%
\bibitem{UA1} C.~Albajar et al. (UA1 Collaboration), Nucl. Phys. B {\bf 335} (1990) 261.
\bibitem{ALICEpPb} B. Abelev et al. (ALICE Collaboration), 
Phys. Rev. Lett. {\textbf{110}} 082302 (2013).
%\cite{Cleymans:2010aa}
%
\end{thebibliography}
\end{document}